\documentclass{ws-p8-50x6-00}
\usepackage{amssymb,bm}

\newcommand{\eeq}{\end{equation}}
\newcommand{\beq}{\begin{equation}}
\newcommand{\ba}{\begin{array}}
\newcommand{\ea}{\end{array}}
\newcommand{\ber}{\begin{eqnarray}}
\newcommand{\eer}{\end{eqnarray}}

\begin{document}

\title{Varying alpha, mass generation and extra dimensions}

\author{Thomas Dent}

\address{{\em Theoretical Physics division,} \\
    {\em University of Ioannina, Ioannina 45110, GREECE} \\
    email: {\tt tdent@auth.gr}}

\maketitle

\abstracts{
\noindent 
We consider variation of coupling strengths and mass ratios in and 
beyond the Standard Model, in the light of various mechanisms of 
mass generation. In four-dimensional unified models, variations in 
charged particle thresholds, light quark masses and the electron 
mass can completely alter the (testable) relation between $\Delta\ln 
\alpha$ and $\Delta \ln \mu$, where $\mu\equiv m_p/m_e$.
In extra-dimensional models where a compactification scale below
the fundamental scale is varying, definite predictions may result
even without unification; we examine some models with
Scherk-Schwarz supersymmetry-breaking.}

\section{Introduction}
\noindent
The recent claim that the fine structure constant $\alpha$ governing QSO
absorption spectra in interstellar gas clouds at redshifts $z=0.5$ to $3$ 
differs from that measured in the laboratory (at 5$\sigma$ level)\cite{Webb03} 
raises many theoretical and experimental issues\cite{Uzan}. 
Prominent among these are the possibility that other fundamental parameters 
of the Standard Model (SM) have also varied, how their variations might be 
bounded or measured and what theories could be tested thereby. Parameters 
that are in principle accessible to astronomical or astrophysical observation 
are, apart from $\alpha$, ratios of particle masses such as $m_q/m_p$ for 
the light quarks or $m_e/m_p$, and dimensionless ratios such as $m_p^2G_F
\propto (m_p/v_H)^2$, where $v_H$ is the Higgs v.\,e.\,v., and $m_p^2G_N
\propto (m_p/M_{\rm Pl})^2$ which measure the strength of weak and 
gravitational interactions respectively. Any variation in the strong 
interaction is equivalent, by dimensional transmutation, to a variation 
in the QCD invariant scale which we denote as $\Lambda_c$ (more properly 
the ratio of $\Lambda_c$ to some other mass scale), or the proton mass. 

The most robust and direct constraints arise from other observations of 
astronomical spectra at comparable redshifts, which are sensitive to the 
parameters $m_p/m_e\equiv \mu$~\cite{Ivanchik} and $g_p$, the gyromagnetic 
ratio of the proton (in addition to $\alpha$)~\cite{CowieS,Murphy01}. The 
second of these observables does not have a well-understood dependence on SM 
parameters \cite{FlambaumS}, so we eliminate it in favour of $\alpha$ and 
$\mu$. Note that the interpretation of such observations depends very little 
on the particular form of the cosmological (space-time) evolution of 
$\alpha$, and is less subject to theoretical uncertainties or degeneracies 
encountered when studying the dependence of nuclear processes or the CMB on 
fundamental parameters.

Then the question we will address is the relation between the variation in
$\alpha$ and that in $\mu$ in theories that predict the values of, or
relations between, the SM parameters. We define the parameter
\beq
 \bar{R} \equiv \frac {\mu^{-1}\Delta\mu} {\alpha^{-1}\Delta\alpha}
\eeq 
which may be compared to observation. Although data constraining $\mu$ are 
scarce and marginally inconsistent, $\bar{R}$ should lie in the range 
$(-10.5,5.5)$ to stand any chance of agreeing with observation. As we will 
see, this already rules out many scenarios.

There are two types of theory where predictions of $\bar{R}$ are possible.
First, those with gauge coupling unification, in which the variation of either 
the unified coupling $g_X$ at the scale $M_X$ (the fundamental scale of the 
theory, often taken as $M_{\rm GUT}$ or $M_{\rm Pl}$) or some ratio of mass 
scales in the theory induces the observed 
variation\cite{C+O,Calmet1,Langacker,Calmet2,us,Dine}. Second, those where 
some or all of the SM fields propagate along (one or more) extra dimensions, 
where the variation of the radius relative to $M_X^{-1}$ induces variation in 
$\alpha$~\cite{PC_S_T}. In this work we include the full effects of thresholds 
and light fermion masses, building on the analysis of \cite{Wetterich02}. These 
effects, although formally subleading, can compete with the zeroth-order term, 
depending on the mechanism of mass generation, and completely alter theoretical 
expectations. For the case of varying extra-dimensional radius, we generalise 
formulae obtained \cite{PC_S_T} for extra-dimensional GUTs \cite{DDG} to cases
without gauge unification. To obtain a meaningful prediction of $\bar{R}$ the 
mechanism of electroweak symmetry-breaking (EWSB) must be accounted for, thus 
we consider models in which this is closely tied to the presence of extra 
dimensions. In general, the inclusion of the effects of mass  generation, 
thresholds and fermion masses can bring the value of $\bar{R}$ closer to the 
range allowed by observation.

\section{High-scale unification and mass generation}
\noindent
The essential feature of GUT-like theories is that gauge couplings are 
determined by a single degree of freedom $g_X$ at the fundamental scale 
$M_X$ and the renormalization group (RG) evolution between $M_X$ and 
observable energies.
We fix notation by quoting (the solution of) the one-loop RG equation
for gauge couplings, including the threshold contribution of a charged field
with decoupling mass $m^{\rm th}(m^{\rm th})$:
\beq \label{eq:2}
 \alpha_i^{-1}(\mu^-) = \alpha_i^{-1}(\mu^+) - \frac{b^-_i}{2\pi} \ln\left(
 \frac{\mu_-}{\mu^+}\right) - \frac{b_i^{\rm th}}{2\pi}
 \ln\left(\frac{m^{\rm th}}{\mu^+}\right)
\eeq
where $b_i$ is negative for asymptotically free groups and 
$b_i^{\rm th}\equiv b_i^+ -b_i^-$, the beta-function coefficient being
$b_i^+$ above the threshold and $b_i^-$ below, with tree-level matching at
$m^{\rm th}$. 

For $\alpha$, we have in general
\beq \label{eq:11}
 \frac{\Delta \alpha}{\alpha} = \sum_{i=1,2}\frac{\alpha}{\alpha_i(\mu^+)}
 \frac{\Delta \alpha_i(\mu^+)}{\alpha_i(\mu^+)}
 + \alpha \sum_{\rm th} \frac{{Q^{\rm th}}^2 f^{\rm th}}{2\pi}
 \Delta \ln \frac{m^{\rm th}}{\mu^+}
\eeq
where $\alpha_1$ = ${g'}^2/4\pi$, the second sum is over all charged fields, 
$f^{\rm th}$ is $2/3$ per chiral (or Majorana) fermion, $1/3$ per complex 
scalar and $11/3$ per vector boson. 
For the light quarks, the QCD scale $\Lambda_c$ provides a dynamical 
cutoff. The first (direct) term gives $\Delta\alpha/\alpha|_{\rm direct} = 
(8\alpha/3\alpha_X) \Delta\ln\alpha_X \simeq 0.47 \Delta\ln\alpha_X$ due to 
the hypercharge normalization in GUTs, where we take the SUSY-GUT value 
$\alpha_X\simeq 1/24$.\footnote{Since single-step non-supersymmetric 
unification is ruled out, in what follows we will take the 
'non-supersymmetric case' to denote the results when superpartner thresholds 
are ignored.}
The remaining threshold contribution is
\bea \label{eq:12}
 \frac{\Delta \alpha}{\alpha}|_{\rm th}
  &=& \frac{\alpha}{2\pi} \left[ \frac{8}{3} \beta_\Lambda \Delta\ln \alpha_X
 + \frac{4}{3} \left(\frac{4}{3} \Delta
 \ln \frac{m_cm_t}{M_X^2} + \frac{1}{3} \Delta\ln \frac{m_b}{M_X}
 + 3\Delta \ln \frac{m_l}{M_X} \right) \right.\nonumber \\
 &-& \left. \frac{21}{3} \Delta \ln \frac{M_W}{M_X} \
  + 8 \Delta\ln \frac{\tilde{m}}{M_X} + \frac{1}{3} \Delta\ln
 \frac{m_H}{M_X} \right] \\
 &=& \frac{\alpha}{2\pi} \left( \frac{8}{3} \beta_\Lambda
 + \beta_v + \frac{25}{3}\beta_S \right) \Delta \ln \alpha_X \label{eq:13}
\eea
where $m_l$ is an averaged lepton mass and 
$\tilde{m}$ and $m_H$ stand for geometric averages over
superpartner and heavy Higgs masses respectively, where such fields exist.
The parameters $\beta_v$ and $\beta_S$ describe the dependence of the
Higgs v.\,e.\,v.\ and superpartner masses on the unified coupling as
\[ \Delta \ln \frac{v_H}{M_X} = \beta_v \Delta \ln \alpha_X,\ 
 \Delta \ln \frac{\tilde{m}}{M_X} = \beta_S \Delta \ln \alpha_X \]
while $\beta_\Lambda$ is defined analogously for the ratio $\Lambda_c/M_X$
(to be derived shortly). We neglect variations in fermion Yukawa couplings.

For the QCD invariant scale $\Lambda_c\equiv Me^{-2\pi/9\alpha_3(M)}$ where 
$m_s<M<m_c$ we have, 
including superpartners (squarks of geometric average mass $m_{\tilde{q}}$ 
and gluinos of mass $m_{\tilde{g}}$), for $\mu^+ > m_t$, $m_{\tilde{q}}$, 
$m_{\tilde{g}}$
\beq \label{eq:4}
 \frac{\Lambda_c}{\mu^+} = e^{-2\pi/9\alpha_3(\mu^+)}
 \left(\frac{m_c m_b m_t}{\mu^{+3}}\right)^{2/27}
 \left(\frac{m_{\tilde{q}} m_{\tilde{g}}}{\mu^{+2}}\right)^{2/9}
\eeq
where for non-supersymmetric theories the last term is to be set to unity.
Thus we find
\beq \label{eq:9}
 \Delta \ln \frac{\Lambda_c}{M_X} \equiv \beta_\Lambda \frac{\Delta 
 \alpha_3(M_X)}{\alpha_3(M_X)}
  = \left(\frac{2\pi}{9\alpha_3(M_X)} + \frac{2}{9}\beta_v
 + \frac{4}{9}\beta_S \right) \frac{\Delta \alpha_3(M_X)}{\alpha_3(M_X)}
\eeq
where we 
set $\beta_S=0$ for a theory without superpartners. 
The threshold terms are of higher order in $\alpha$ or $\alpha_3$, thus 
formally they ought to be grouped with the power-law correction to 
$\Lambda_c$ from two-loop running, which we have neglected. However, when 
the Higgs v.\,e.\,v.\ is varying rapidly compared to $\alpha$, as is 
generic in theories with high fundamental scale~\cite{C+O,Langacker,us}, 
the threshold terms can become dominant, in contrast to the two-loop term, 
which is model-independent. 

The values of $\beta_v$ and $\beta_S$ are determined by the mechanisms of 
mass generation and SUSY-breaking. Technicolour and radiative EWSB with 
hidden-sector SUSY-breaking have in common the generation of an 
exponentially small scale by strong running in an asymptotically-free gauge
group: 
\beq \label{eq:6}
 \frac{v_H}{M_X} = k \alpha_X^n e^{-2\pi m/b_h\alpha_X}
\eeq
where $k$ is a numerical constant and $n$ parameterises a possible power-law 
dependence. Here we take the most natural case when the coupling of the
the ``hidden'' gauge group responsible for generating the electroweak scale 
$\alpha_H$ is unified with the SM gauge couplings at the scale $M_X$. Our 
estimates are valid more generally if the variation in $\alpha_H^{-1}(M_X)$ 
equals that in $\alpha_X^{-1}$.
In the case of theories without fundamental Higgs, this equation simply 
parameterises whatever condensate breaks SU$(2)\times$U$(1)$. Thus,
\beq \label{eq:7}
 \frac{M_X}{v_H} \Delta \frac{v_H}{M_X} \simeq \left(n +
 \frac{2\pi m}{b_h\alpha_X}\right) \frac{\Delta \alpha_X}{\alpha_X}.
\eeq
Assuming that $k$ and $n$ are order 1, we neglect (the logarithm of) 
their contribution to Eq.~(\ref{eq:6}) and find
\[ \frac{2\pi m}{b_h\alpha_X} \simeq \ln[M_X/v_H \simeq
 (2\times 10^{16})/(2\times 10^2)] \simeq 32
\]
where we identified $M_X$ with the SUSY-GUT value $2.4\times 10^{16}\,$GeV.
Thus the RHS of Eq.~(\ref{eq:7}) is approximately $(n + 32) \Delta \ln
\alpha_X$, consistent with neglecting $n$ of order 1. If $M_X$ is the 
heterotic string scale $M_X\sim 4\times 10^{17}\,$GeV the prefactor is 
$(n + 35)$, thus the uncertainty introduced by $n$ is small and we estimate 
in general $\beta_v =(34\pm 2)$.
Similarly, SUSY-breaking masses vary as
\beq \label{eq:8}
\frac{\tilde{m}}{M_X} = k' \alpha_X^{n'} e^{-2\pi m/b_h\alpha_X}
\eeq
where $m_{\tilde{q}}$, $m_{\tilde{g}}$ {\em etc.}\/\ vary as $\tilde{m}$,
thus their variation can also be estimated as $\beta_S =34\pm 2$ given that
superpartner masses are around the electroweak scale.

Radiative EWSB may, however, depend sensitively on a combination of the 
top Yukawa coupling and ratios of SUSY-breaking soft mass terms 
\cite{RobertsRoss,C+O}, which are exceedingly model-dependent. 
To cover potential uncertainties we allow the Higgs v.\,e.\,v.\ to have a 
different fractional variation compared to the superpartner masses, thus we 
allow $\beta_v\neq \beta_S$ in general. Since there is also the possibility 
that the variation of the ``hidden'' gauge kinetic function 
$\alpha_H(M_X)^{-1}$ may differ from that of $\alpha_X^{-1}$, we keep 
$\beta_S$ and $\beta_v$ as free parameters in the calculation and take the 
values $\beta_v= \beta_S= 34\pm 2$ as illustrative only.
For these values, using the dependence of $\Lambda_c$ from Eq.~(\ref{eq:9}) 
in Eq.~(\ref{eq:12}) we find 
\[ \Delta\ln \alpha|_{\rm th} \simeq
(0.11\pm 0.01) \Delta\ln \alpha_X\ \mbox{[non-SUSY]},\
(0.49 \pm 0.03) \Delta\ln \alpha_X\ \mbox{[SUSY]}
\]
which is a non-negligible correction to the direct contribution 
$ 0.47 \Delta\ln \alpha_X$. 

Now if $m_p$ is well-approximated by a constant times $\Lambda_c$, we find 
\beq \label{eq:10}
\frac{\Delta \mu}{\mu}
 = \left(\beta_\Lambda - \beta_v \right) \frac{\Delta \alpha_X}{\alpha_X}
 = (-10 \pm {\rm few}) \frac{\Delta \alpha_X}{\alpha_X} \ \mbox{[non-SUSY],}\ \ 
 (5 \pm {\rm few}) \frac{\Delta \alpha_X}{\alpha_X} \ {\rm[SUSY]}.
\eeq
If the electron Yukawa coupling also varies, due to some dynamics of flavour 
structure, these values may be changed to, for example, $-13$ (non-SUSY) or 
$+2$ (SUSY) if we have $y_e\propto \alpha_X^3$. These values differ
considerably from the cases where Higgs v.\,e.\,v.\ and superpartner masses 
are fixed relative to $M_X$, giving $\Delta\ln \mu \simeq 17 \Delta\ln 
\alpha_X$, or where $v_H$ varies as in (\ref{eq:7}) but the resulting effects 
on and of QCD thresholds are neglected, giving $\Delta\ln \mu/\mu \simeq -17 
\Delta\ln \alpha_X$. The values of $\bar{R}$ following from Eq.~\ref{eq:10} 
are $-17\pm 3$ (non-SUSY) and $4\pm 3$(SUSY). 

One may also include the effect of varying $m_u$, $m_d$ and $m_s$ relative 
to $\Lambda_c$ on the proton mass: we find \cite{me03}
\beq
\Delta\ln \frac{m_p}{M} 
  \approx 0.78 \Delta\ln \frac{\Lambda_c}{M} 
 + (0.12\pm 0.1) \Delta \ln\frac{m_s}{M}
 + 0.048 \sum_{u,d} \Delta \ln\frac{m_q}{M} 
\label{eq:14}
\eeq
where the strange term introduces the largest uncertainty, the other terms 
having errors of order $10\%$. Thus our final estimate is
\beq
\bar{R} \approx
 \frac{0.54 \alpha_X^{-1} + (-0.61\pm 0.12)\beta_v +0.35\beta_S }
 {0.022 \alpha_X^{-1} + 0.0018\beta_v + 0.011\beta_S }
\eeq
Without variation in the electroweak and superpartner scales, setting
$\beta_v= \beta_S= 0$ we find the rather model-independent result
$\bar{R}= 25\pm ({\rm few})$. This differs from the expectation
$\bar{R}\simeq 36$ 
obtained by neglecting the effects of light quark masses on $\Lambda_c$ and 
the effect of varying $\Lambda_c/M_X$ on $\alpha$. If we 
set
$\beta_v\simeq 34$
and $\beta_S=0$, we obtain
$\bar{R}= -13 \pm 7$. 
Finally for $\beta_S= \beta_v= 34\pm2$ we find
$\bar{R}= 4 \pm 5$. 

\section{Varying extra-dimensional radii}
In a model which is described by ($4+\delta$)-dimensional field theory
over some range of energies, there is a hierarchy between the (inverse)
compactification radius or radii $R_i^{-1}$ and the ultraviolet cutoff of
the higher-dimensional theory $\Lambda_D$, where $D=4+\delta$ 
\cite{Antoniadis,DDG}. This makes
it a well-defined problem to compute the effect of varying radius on the
4D low-energy theory, since one can take physics above the scale
$R_i^{-1}$ to be unchanged by such variation (apart from the change in
masses of Kaluza-Klein modes), consistent with decoupling.
The varying dimensionless quantity is the radius normalized to the 
cutoff of the extra-dimensional field theory, $R\Lambda_D$. By dimensional 
reduction at the energy scale $M_\delta\equiv R^{-1}$ we find the simple 
formulae
\[
\Delta\ln \alpha_i(M_\delta) = \delta_i 
\Delta\ln \frac{M_\delta}{\Lambda_D},\ 
\Delta\ln y(M_\delta) = \delta_i \left(\frac{p_y}{2}+c\right)
 \Delta\ln \frac{M_\delta}{\Lambda_D}
\]
which are good approximations in the perturbative regime \cite{me03}, 
where $y$ is a Yukawa coupling for an operator with $p_y=0,1,2,3$ fields 
propagating in extra dimension and $(3-p_y)$ fields localized ``on the brane'' 
and $\delta_i$ is the number of extra dimensions around which the gauge or 
matter fields propagate. For simplicity we will take one varying dimension. 
The integer $c$ takes the value $0$ for operators localized in the extra 
dimension 
and $1$ for a bulk coupling when $p_y=3$. 
For a supersymmetric Higgs mass term $\mu_SH_1H_2$ localized on the brane 
with Higgs in the bulk, $\mu_S$ is directly proportional to $M_\delta$, 
$\mu_S/M_\delta= \mbox{constant}$; for a bulk mu-term or for localized 
Higgses $\mu_S$ arises from a mass term in the $D=5$ theory and varies as 
$\mu_S/M_\delta\propto (M_\delta/\Lambda_D)^{-1}$.

If we now take the Higgs v.\,e.\,v.\ and masses of SM fermions (and 
superpartners, if any) to vary proportional to $M_\delta$, which is expected 
to be a reasonable approximation for Scherk-Schwarz symmetry-breaking, we find 
\bea
\Delta \ln \alpha &=& \left(\delta_Y\cos^2 \theta_W + \delta_2\sin^2 \theta_W
 + \frac{\alpha(M_Z)}{2\pi}(\delta_Y b_Y+\delta_2 b_2)
 \ln \frac{M_Z}{M_\delta} \right) \Delta \ln \frac{M_{\delta}}{\Lambda_D},
 \nonumber \\
\Delta \ln \mu &=& \frac{2\pi\delta_3}{9}\left(\alpha^{-1}_3(M_Z) +
 \frac{b_3}{2\pi}\ln \frac{M_Z}{M_\delta}\right) \Delta\ln
 \frac{M_{\delta}}{\Lambda_D}
\eea
thus without superpartners we have
\beq \label{eq:21}
\bar{R} \approx \delta_3\left(5.9 -\frac{7}{9}\ln \frac{M_Z}{M_\delta}
 \right)
 \left[ \delta_Y\cos^2 \theta_W + \delta_2\sin^2 \theta_W
 + \frac{1}{804}\left(\frac{41}{6} \delta_Y -
 \frac{19}{6} \delta_2\right)\ln \frac{M_Z}{M_\delta} \right]^{-1}
\eeq
and with superpartners (in the case $M_\delta \gg \tilde{m}$)
\beq \label{eq:22}
\bar{R} \approx \delta_3\left(5.9 -\frac{3}{9}\ln \frac{M_Z}{M_\delta}
 \right)
 \left[ \delta_Y\cos^2 \theta_W + \delta_2\sin^2 \theta_W
 + \frac{1}{804}\left(11 \delta_Y +
 \delta_2\right)\ln \frac{M_Z}{M_\delta} \right]^{-1}.
\eeq
Enforcing $\delta_Y=\delta_2=\delta_3$ as required by an
extra-dimensional GUT we obtain
\beq \label{eq:23}
\bar{R} \approx \frac{ 5.9 - \frac{7}{9}\ln (M_Z/M_\delta) }
 { 1+\frac{1}{804}\cdot\frac{11}{3} \ln (M_Z/M_\delta) }\ \mbox{[non-SUSY]},\
\frac{ 5.9- \frac{3}{9}\ln (M_Z/M_\delta) }
 { 1+\frac{1}{804}\cdot 12 \ln (M_Z/M_\delta) }\ \mbox{[SUSY]}.
\eeq
If we use the more detailed treatment of the proton mass Eq.~(\ref{eq:14})
then the value of $\bar{R}$ is simply reduced by a factor $0.78$.
Given Eq.~(\ref{eq:23}), in the small radius limit $M_\delta\rightarrow 
2.4\times 10^{16}\,$GeV we recover the SUSY-GUT expectation $\bar{R}\approx 
33$. In the opposite limit where $M_\delta$ approaches $M_Z$ we would obtain 
$\bar{R}\gtrsim 6$. This suppression of $\bar{R}$ is consistent with 
\cite{PC_S_T}.

In more general ``brane world'' models, one has freedom to choose the 
integers $\delta_i$ and $p_y$. Clearly if $\delta_3=0$ we have 
$\Delta\ln \mu=0$, up to effects (so far neglected) of thresholds
below $M_\delta$ and variation in $m_e/M_\delta$. The strong force does 
not ``feel'' the variation because it is not propagating round the varying 
dimension. If one wishes to fit the value $\bar{R}=0$ this choice of 
$\delta_3$ is obvious, if somewhat arbitrary in the absence of a concrete 
model.

Above, we neglected any variation in the electron mass $m_e/M_\delta$. But 
for small values of $\bar{R}$ even a mild power-law dependence of 
$m_e/M_\delta$ may be significant. We considered \cite{me03} some concrete 
models of EWSB in extra dimensions \cite{Quiros,Nomurawk}: the model of 
\cite{Quiros} does not lead to any definite prediction due to the arbitrary 
choice of the value of $\mu_S$ and the sensitivity of the variation of 
$v_H/M_\delta$ to this parameter. The model of \cite{Nomurawk} predicts the 
dependence $m_e/M_\delta \simeq \mbox{constant}\times (M_\delta/\Lambda_D)$, 
which simply subtracts $1$ from the denominator of Eq.~\ref{eq:23}. This 
feeble dependence justifies neglecting the variation of other fermion masses. 
Taking into account the compactification scale required by the model we find 
$\bar{R} = 6.3 \pm 1$.

\section{Conclusions}
If the finding of a nonzero variation in $\alpha$ persists, improved 
constraints on $\mu$ may be a powerful tool to discriminate between models of 
physics beyond the SM. We showed that it is important to include effects of 
varying mass ratios which were previously neglected in models with high-scale
unification. We also showed that it is possible to obtain predictions without
unification if the observed variation is due to a varying extra dimension.


\section*{Acknowledgements}
Work supported by the EU Fifth Framework Network `Across the Energy Frontier' 
(HPRN-CT-2000-00148).


\begin{thebibliography}{99}

\bibitem{Webb03}
M.\,T.~Murphy, J.\,K.~Webb and V.\,V.~Flambaum,
Mon.\ Not.\ Roy.\ Astron.\ Soc.\  {\bf 345} (2003) 609
[astro-ph/0306483];
J.\,K.~Webb, M.\,T.~Murphy, V.\,V.~Flambaum and S.\,J.~Curran,
Astrophys.\ J.\ Supp.\  {\bf 283} (2003) 565
[astro-ph/0210531],
astro-ph/0209488.

\bibitem{Uzan}
J.~P.~Uzan,
Rev.\ Mod.\ Phys.\  {\bf 75} (2003) 403
[hep-ph/0205340].

\bibitem{Ivanchik}
A.\,V.~Ivanchik {\it et al.},
Astron.\ Lett.\  {\bf 28} (2002) 423
[astro-ph/0112323].

\bibitem{CowieS}
L.\,L.~Cowie and A.~Songaila,
Astrophys.\ J.\  {\bf 453} (1995) 596.

\bibitem{Murphy01}
M.\,T.~Murphy {\it et al.},
Mon.\ Not.\ Roy.\ Astron.\ Soc.\  {\bf 327} (2001) 1244.

\bibitem{FlambaumS}
V.\,V.~Flambaum and E.\,V.~Shuryak,
Phys.\ Rev.\ D {\bf 65} (2002) 103503;
V.\,F.~Dmitriev and V.\,V.~Flambaum,
Phys.\ Rev.\ D {\bf 67} (2003) 063513.

\bibitem{C+O}
B.\,A.~Campbell and K.\,A.~Olive,
Phys.\ Lett.\ B {\bf 345} (1995) 429.

\bibitem{Langacker}
P.~Langacker, G.~Segr{\' e} and M.\,J.~Strassler,
Phys.\ Lett.\ B {\bf 528} (2002) 121.

\bibitem{Calmet1}
X.~Calmet and H.~Fritzsch,
Eur.\ Phys.\ J.\ C {\bf 24} (2002) 639.

\bibitem{Calmet2}
X.~Calmet and H.~Fritzsch,
Phys.\ Lett.\ B {\bf 540} (2002) 173.

\bibitem{us}
T.~Dent and M.~Fairbairn,
Nucl.\ Phys.\ B {\bf 653} (2003) 256.

\bibitem{Dine}
M.~Dine, Y.~Nir, G.~Raz and T.~Volansky,
Phys.\ Rev.\ D {\bf 67} (2003) 015009.

\bibitem{PC_S_T}
F.~Paccetti Correia {\em et al.},
hep-ph/0211122.

\bibitem{Wetterich02}
C.~Wetterich,
JCAP {\bf 0310} (2003) 002
[arXiv:hep-ph/0203266].

\bibitem{DDG}
K.\,R.~Dienes, E.~Dudas and T.~Gherghetta,
Nucl.\ Phys.\ B {\bf 537} (1999) 47
[hep-ph/9806292],
Phys.\ Lett.\ B {\bf 436} (1998) 55
[hep-ph/9803466].

\bibitem{me03}
T.~Dent,
hep-ph/0305026.

\bibitem{RobertsRoss}
G.\,G.~Ross and R.\,G.~Roberts,
Nucl.\ Phys.\ B {\bf 377} (1992) 571.

\bibitem{Antoniadis}
I.~Antoniadis,
Phys.\ Lett.\ B {\bf 246} (1990) 377.

\bibitem{Quiros}
A.~Delgado and M.~Quiros,
Nucl.\ Phys.\ B {\bf 607} (2001) 99.

\bibitem{Nomurawk}
R.~Barbieri, L.\,J.~Hall and Y.~Nomura,
Phys.\ Rev.\ D {\bf 63} (2001) 105007.


\end{thebibliography}
\end{document}